\documentclass[showpacs,preprintnumbers,amsmath,amssymb,prb,twocolumn]{revtex4}

\usepackage{graphicx}
\usepackage{dcolumn}
\usepackage{bm}
\def\comment#1{}

\begin{document}

\title{Critical dynamics, duality, and the
exact dynamic exponent in extreme type II superconductors}
\author{Flavio S. Nogueira}
\email{nogueira@physik.fu-berlin.de}
\affiliation{Institut f{\"u}r Theoretische Physik,
Freie Universit{\"a}t Berlin, Arnimallee 14, D-14195 Berlin, Germany}

\author{Dirk Manske}
\email{d.manske@fkf.mpg.de}
\affiliation{Max-Planck-Institut f{\"u}r Festk{\"o}rperforschung,
Heisenbergstr. 1, D-70569 Stuttgart, Germany}

\date{Received \today}

\begin{abstract}
The critical dynamics of superconductors is studied
using renormalization group and duality arguments.
We show that in extreme type II superconductors
the dynamic critical exponent is given exactly by $z=3/2$.
This result does not rely on the widely used models
of critical dynamics. Instead, it is shown that $z=3/2$ follows
from the duality between the extreme type II superconductor and a
model with a critically fluctuating gauge field. Our result is
in agreement with Monte Carlo simulations and at least one experiment.
\end{abstract}

\maketitle

\section{Introduction}

The high-$T_c$ cuprate superconductors have very
large values of the Ginzburg parameter $\kappa$. In a
material like YBa$_2$Cu$_3$O$_{7-\delta}$ (YBCO) at optimal doping, we have
$\kappa\sim 100$. It is therefore a good approximation
to assume that these materials are extreme type II
superconductors. By extreme type II superconductor we
mean $\kappa\to\infty$. In such a regime
it is expected that the static critical
properties at zero external field are the same as
in superfluid $^4$He. \cite{Schneider} This expectation
is confirmed by experiments measuring specific heat
\cite{Salamon} and penetration depth \cite{Kamal} in
bulk samples of YBCO. Thus, as far as static critical
phenomena are concerned, there is no doubt that
the critical behavior of bulk YBCO is governed by
the three-dimensional $XY$ universality class.
On the basis of such a consensus, we might also expect
that the dynamical universality class is the same
as in superfluid $^4$He. Unfortunately, we are far from
reaching a consensus on the dynamical universality
class of YBCO, or more generally, of any high-$T_c$ cuprate.
What is interesting here is that the lack of consensus comes
both from the theoretical and experimental sides. The
theoretical debate tries to establish whether the
dynamical universality class corresponds to model A or model F
dynamics. \cite{Hohenberg} Model A dynamics is purely
relaxational and gives
the value $z\approx 2.015$ for the
dynamic critical exponent in three dimensions.
Model F, on the other hand, features a conserved
density coupled to a spin-wave mode and gives the {\it exact} value
$z=d/2$ for $d\in(2,4]$. For the dynamical universality
class of extreme type II superconductors, Monte Carlo simulations
give $z\approx 3/2$, \cite{Weber,Lidmar} which would be
consistent with model F dynamics
and therefore with superfluid $^4$He dynamical universality class.
However, in a recent letter Agi and Goldenfeld \cite{Golden}
claim that the correct result from a lattice model should be
instead $z\approx 2$, i.e., consistent with model A dynamics.
From the experimental side it is found either large values
for $z$, typically in the range $z\sim 2.3-3.0$, \cite{Booth}
either values consistent with model F dynamics. \cite{Kim}
The arguments of Ref. \onlinecite{Golden} were further discussed
in two recent comments. \cite{comment}

In principle model F dynamics is not compatible with
superconductor critical dynamics, since screening effects
tend to suppress the spin-wave mode.
\cite{Fisher,Dorsey} 
However, model A is also
inappropriate to study the critical dynamics
in superconductors. Indeed, model A does not give
a gauge-independent result for $z$ in the magnetic critical fluctuation
(MCF)  regime. A technically
correct analysis should consider the extreme type II regime as
a limit of the full MCF regime which has a genuine local gauge
symmetry. It turns out that in a gauge-invariant theory
the only operators with a nonzero expectation value are
the gauge-invariant ones. \cite{Elitzur} For instance, the AC conductivity
is such a gauge-invariant quantity. Therefore, its scaling
behavior is necessarily gauge-independent. Since evaluation
of $z$ through model A in the MCF regime gives a gauge-dependent
result, we conclude that model A is also not the right option.

In this paper we use renormalization group (RG) and duality
arguments to determine the dynamical universality class of
extreme type II superconductors. We will establish that
in three dimensions $z=3/2$ {\it exactly}. This result will
be obtained by combining exact scaling arguments with
exact duality results. While the scaling arguments are
generally valid in $d\in(2,4]$, the duality arguments will
be only valid at $d=3$. The result will be obtained through
the following strategy. In Section II we use the RG to obtain
exact scaling relations for the penetration depth and
for the AC conductivity. This will be done in both the
$XY$ (extreme type II limit)
and MCF regime. While much
of the steps in this part of derivation are known, it is
important to review this approach here to emphasize the
intimate relationship between the scaling of the penetration
depth and the one of the AC conductivity. The important
point is that scaling behavior in the $XY$ regime is different
from the one in the MCF regime. In Section III 
we discuss critical dynamics from 
the perspective of the exact duality between an extreme type II
superconductor at zero field and a model exactly
equivalent to the Ginzburg-Landau (GL) model in the
MCF regime. Indeed, the dual model features a
fluctuating vector potential coupled to a complex field, 
dubbed disorder parameter field as opposed to the order
parameter field of the original model. The derived results
for the GL model in the MCF regime will then be applied to
the dual model and duality relations between the currents
will be used to establish that $z=3/2$.

\section{Scaling and dynamics in the Ginzburg-Landau model}

In order to make the paper self-contained, 
we review in this Section the basic scaling properties of the GL model in both 
the static and dynamic critical regimes. 

\subsection{Static scaling behavior}

Let us consider the {\it bare} Hamiltonian of the GL model,

\begin{equation}
\label{GL}
{\cal H}=\frac{1}{2}(\mbox{\boldmath $\nabla$}\times{\bf A}_0)^2+
|(\nabla-iq_0{\bf A}_0)\psi_0|^2+\mu_0^2|\psi_0|^2
+\frac{u_0}{2}|\psi_0|^4,
\end{equation}
where $q_0=2e_0$ is the charge of the Cooper pair and
$\mu_0^2\propto \tau$, with $\tau=(T-T_c)/T_c$. In our notation
the zero subindex denotes bare quantities while
in renormalized quantities the zero subindices are absent.
The bare Ginzburg parameter is $\kappa_0=\lambda_0/\xi_0=(u_0/2q_0^2)^{1/2}$,
where $\lambda_0$ and $\xi_0\equiv\mu_0^{-1}$
are the bare penetration depth and correlation length,
respectively. We can rewrite the above Hamiltonian in terms
of renormalized quantities as follows:

\begin{equation}
\label{renormGL}
{\cal H}=\frac{Z_A}{2}(\mbox{\boldmath $\nabla$}\times{\bf A})^2+
Z_\psi|(\nabla-iq{\bf A})\psi|^2+Z_\mu \mu^2|\psi|^2
+\frac{Z_u u}{2}|\psi|^4.
\end{equation}
The renormalized correlation length is given by
$\xi=\mu^{-1}$. From the Ward identities it follows
that the renormalized charge squared is given by
$q^2=Z_A q_0^2$. The dimensionless
couplings are $f\equiv\mu^{d-4}q^2$ and
$g\equiv\mu^{d-4}Z_\psi^2 u_0/Z_u$. The fixed
point structure is well known but cannot be
completely obtained by perturbative means. Fixed
points associated to nonzero charge, $f_*\neq 0$, are
non-perturbative but their existence in the flow diagram
is now well established. 
\cite{Dasgupta,KleinertLNC,
Kiometzis,Berg,Folk,Herbut,Olsson,deCalan,Hove,KleinNog}
The infrared stable charged fixed point governs the MCF
regime while the extreme type II or $XY$ regime is governed
by the {\it uncharged} $XY$ fixed point \cite{KleiNogHol}. 

The renormalized Ginzburg parameter
is given by $\kappa=\mu/\mu_A$, where $\mu_A=\lambda^{-1}$
is the renormalized vector potential mass generated by the Anderson-Higgs 
mechanism.  
Due to the Ward
identities this can also be written as $\kappa=(g/2f)^{1/2}$, and
therefore the renormalized Ginzburg parameter has the same form
as the bare one, with the bare coupling constants replaced by the 
renormalized ones. 
From this it can be shown that the following 
{\it exact} evolution equation
for the renormalized vector potential mass holds: \cite{deCalan}
\begin{equation}
\label{flowmA}
\mu\frac{\partial\mu_A^2}{\partial\mu}=
\left(d-2+\gamma_A-\frac{\beta_g}{g}\right)\mu_A^2,
\end{equation}
where $\gamma_A(f,g)\equiv\mu\partial\ln Z_A/\partial\mu$ and 
$\beta_g\equiv\mu\partial g/\partial\mu$. 
Eq. (\ref{flowmA}) implies that near the phase
transition 
\begin{equation}
\mu_A\sim\mu^{(d-2+\eta_A)/2}, 
\end{equation}
where $\eta_A=\gamma_A^*$ is the anomalous dimension of the vector 
potential. In the MCF regime $\eta_A=4-d$, \cite{Berg,Herbut,Hove} and 
we obtain that  $\nu'=\nu$, \cite{Herbut} where $\nu'$ and 
$\nu$ are the penetration depth and correlation length 
exponents, respectively. This result was confirmed
by Monte Carlo simulations. \cite{Olsson}
In the $XY$ regime, on the other hand, the penetration depth exponent is
given by \cite{Fisher}

\begin{equation}
\nu'=\frac{\nu(d-2)}{2},
\end{equation} 
where $\nu$ is
the correlation length exponent. At $d=3$
we have $\nu\approx 2/3$ and $\nu'\approx 1/3$. The
result $\nu'\approx 1/3$ is confirmed
experimentally in high quality single crystals
of YBCO. \cite{Kamal} Note that in the $XY$ regime
$\kappa_0\to\infty$ while $\kappa\to 0$. \cite{Fisher}

\subsection{Dynamic scaling behavior}

The AC conductivity is given by

\begin{equation}
\label{ACconduc}
\sigma(\omega)=\frac{q^2 K(-i\omega)}{-i\omega},
\end{equation}
where $K(-i\omega)$ is obtained from the current-current correlation
function at zero momentum, \cite{Halperin} i.e.,
$K(-i\omega)=\lim_{|\bf p|\to 0}K(-i\omega,{\bf p})$ where
$K(-i\omega,{\bf p})=\sum_\mu K_{\mu\mu}(-i\omega,{\bf p})$ with

\begin{eqnarray}
\label{curr-correl}
K_{\mu\nu}(-i\omega,{\bf p})=
\langle|\psi|^2\rangle\delta_{\mu\nu}-\frac{1}{q^2}
\langle\hat{J}_\mu(\omega,{\bf p})\cdot\hat{J}_\nu(-\omega,{\bf p})\rangle,
\end{eqnarray}
and $\hat{J}_\mu(\omega,{\bf p})$ is
the Fourier transform of the
superconducting current 

\begin{equation}
J_\mu=-iq(\psi^*\partial_\mu\psi
-\psi\partial_\mu\psi^*)-2q^2|\psi|^2 A_\mu. 
\end{equation}
The function
$K_{\mu\nu}(-i\omega,{\bf p})$ is purely transverse. \cite{Halperin}
The superfluid density
$\rho_s$ is given by $\lim_{\omega\to 0}K(-i\omega)$ and thus, in virtue of
the Josephson relation, we obtain 

\begin{equation}
\lim_{\mu\to 0}K(-i\omega)\sim(-i\omega)^{(d-2)/z}.
\end{equation}
Since $q^2\sim\mu^{\eta_A}$, we obtain from
Eq. (\ref{ACconduc}) the behavior

\begin{equation}
\label{ACconduccrit}
\sigma(\omega)|_{T=T_c}\sim(-i\omega)^{(d-2-z+\eta_A)/z}.
\end{equation}
Therefore, in the $XY$ universality class we have
\cite{Fisher,Dorsey}

\begin{equation}
\label{ACconduccritXY}
\sigma(\omega)|_{T=T_c}\sim(-i\omega)^{(d-2-z)/z}, 
\end{equation}
while in the MCF regime
we obtain

\begin{equation}
\label{ACconduccritMCF}
\sigma(\omega)|_{T=T_c}\sim(-i\omega)^{(2-z)/z}.
\end{equation}

From Eq. (\ref{ACconduc}) we obtain that below $T_c$ the low frequency behavior of the AC conductivity
is

\begin{equation}
\label{ACconduc1}
\sigma(\omega)\approx\frac{\mu_A^2}{-i\omega}.
\end{equation}
In Ref. \cite{Fisher} this is written symply as
$\sigma(\omega)\approx\rho_s/(-i\omega)$, i.e.,
the charge is not shown explicitly
since only the $XY$ regime was considered and in this case
the charge does not fluctuate. Taking the low
frequency behavior (\ref{ACconduc1}) into account, we
obtain from Eqs. (\ref{flowmA}) and
(\ref{ACconduc}),

\begin{equation}
\label{flowconduc}
\mu\frac{\partial\sigma(\omega)}{\partial\mu}
\approx\left(d-2-z+\gamma_A-\frac{\beta_g}{g}\right)
\sigma(\omega),
\end{equation}
which implies that near
the phase transition,

\begin{equation}
\label{conducscal}
\sigma(\omega)\sim\mu^{d-2-z+\eta_A}.
\end{equation}
Since in the $XY$ universality class $\eta_A=0$, we recover
from Eq. (\ref{conducscal}) the well known scaling \cite{Fisher}

\begin{equation}
\label{conducscal0}
\sigma(\omega)\sim\mu^{d-2-z}.
\end{equation} 
Note that
Fisher {\it et al.} \cite{Fisher} need to assume
the Josephson relation $\rho_s\sim\mu^{d-2}$ to derive
the $XY$ scaling of the AC conductivity. Within our approach
the more general scaling relation (\ref{conducscal})
follows from Eq. (\ref{flowmA}) and the $XY$ scaling
emerges as a particular case. In the MCF regime we
obtain \cite{Mou,Nogueira}

\begin{equation}
\label{conducscal1}
\sigma(\omega)\sim\mu^{2-z}.
\end{equation}

\section{Duality and critical dynamics}

\subsection{Duality and disorder field theory}

In the extreme type II limit
$\kappa_0\to\infty$ and we have essentially a superfluid
model at zero field, i.e., the corresponding Hamiltonian
is the same as in Eq. (\ref{GL}) with ${\bf A}_0=0$.
The lattice version of this model in the London limit is
exactly dual to the so called ``frozen'' superconductor
\cite{Peskin,Neuhaus}. Starting from the Villain form
of the $XY$ model we obtain, after dualizing it, the
following lattice model Hamiltonian:

\begin{equation}
\label{latdual}
H=\sum_l\left[\frac{1}{2K}(\mbox{\boldmath $\nabla$}\times{\bf h}_l)^2
-2\pi i ~{\bf M}_l\cdot{\bf h}_l\right],
\end{equation}
where $K$ is the bare superfluid stiffness, $a_{l\mu}\in(-\infty,\infty)$
and $M_{l\mu}$ is an integer link variable satisfying
the constraint $\mbox{\boldmath $\nabla$}\cdot{\bf M}_l=0$. The lattice
derivative is defined as usual,
$\nabla_\mu f_l\equiv f_{l+\hat{\mu}}-f_l$.
The link variables play the role of vortex currents and the
zero lattice divergence constraint means that only closed
vortex loops should be taken into account.
Integration over
${\bf a}_l$ gives a long range interaction between the link variables.
The link variables will interact through a potential
$V({\bf r}_l-{\bf r}_{m})$ behaving at large distances like
$V({\bf r}_l-{\bf r}_{m})\sim 1/|{\bf r}_l-{\bf r}_{m}|$.
At short distances the potential is divergent. This
short distance divergence can be regularized by adding to
the Hamiltonian (\ref{latdual}) a core energy term
$(\varepsilon_0/2)\sum_l{\bf M}_l^2$. Writing the constraint
$\mbox{\boldmath $\nabla$}
\cdot{\bf M}_l=0$ using the integral representation of
the Kronecker delta and performing the sum over
${\bf M}_l$ using the Poisson formula we arrive at the
Hamiltonian
\begin{eqnarray}
\label{latdual1}
H&=&\sum_l\left[\frac{1}{2\varepsilon_0}
(\mbox{\boldmath $\nabla$}\theta_l-2\pi{\bf N}_{l}-2\pi\sqrt{K} ~
{\bf h}_{l})^2
\right.\nonumber\\&+&\left.
\frac{1}{2}(\mbox{\boldmath $\nabla$}\times{\bf h}_l)^2\right],
\end{eqnarray}
where we have rescaled ${\bf h}_l$. The dual lattice Hamiltonian
(\ref{latdual1}) has exactly the same form as the Hamiltonian for
a Villain lattice superconductor. Note that $2\pi\sqrt{K}$ plays the
role of the charge.
The sum over the integers ${\bf N}_i$ can be converted into a
disorder field theory (DFT) \cite{KNC}
which has
precisely the same form as
the original
 GL model in Eq.~(\ref{GL}), except that the
physical properties of the fields have changed.
The electromagnetic vector potential ${\bf A}_0$ is replaced
by the gauge field ${\bf h}_0$ describing vortices,
 and the charge
$q_0$ becomes the Biot-Savart-type
coupling strength
between vortices $2\pi\sqrt{\rho_s^0}$, where $\rho_s^0$ is the
bare superfluid density. An important aspect of duality is 
that the disorder field theory that comes out of it has 
an ``inverted'' temperature axis, \cite{Dasgupta} i.e., 
the broken symmetry phase of the disorder field theory 
corresponds to the symmetric phase of the original theory 
and vice-versa. This is actually the meaning of the expression 
``disorder field'' used in this paper: instead of having 
an {\it order parameter} as in the original model, the dual 
model has a {\it disorder parameter}. Both models describe 
the same physics and have of course the {\it same} critical temperature 
$T_c$. Since the Ward identities for either 
theory imply that the critical singularities are the same 
irrespective of whether $T_c$ is approached from above or from below, 
we can use the {\it same} scale $\mu$ to study the scaling behavior 
in both models. \cite{deCalan,Tesanovic} 
The critical exponents
$\alpha$ and $\nu$ of the dual model are the same as in the
original model. This is because the dual model give
the {\it same} free energy of the original model, up to non-singular
terms. Therefore, the exponent appearing in the scaling of the
singular part of the free energy, $\alpha$,
is the same in both models. The hyperscaling relation then implies
that $\nu$ is also the same in both models.

The continuum dual model for superconductors was introduced in Ref. 
\onlinecite{Kiometzis} and further discussed in Ref. \onlinecite{Herbut1} 
(see also Ref. \onlinecite{Son}). 
Such a continuum dual theory represents a generalization of the London model. 
In the extreme type II limit and zero external magnetic field, it is obtained 
from the continuum limit of Eq. (\ref{latdual1}),

\begin{equation}
\label{contdual}
\tilde{\cal H}=\frac{1}{2}(\mbox{\boldmath $\nabla$}\times{\bf h}_0)^2+
|(\mbox{\boldmath $\nabla$}-i\tilde{q}_0{\bf h}_0)\phi_0|^2+m_0^2|\phi_0|^2
+\frac{v_0}{2}|\phi_0|^4,
\end{equation}
where the {\it dual bare charge} is given by 

\begin{equation}
\tilde{q}_0=\frac{2\pi\mu_{A,0}}{q}=2\pi\sqrt{\rho_s^0}.
\end{equation}

The Hamiltonian (\ref{contdual}) has the same form as the GL Hamiltonian 
in Eq. (\ref{GL}). In Eq. (\ref{contdual}) the gauge field ${\bf h}_0$ 
is minimally coupled to the already mentioned {\it disorder field} 
$\phi_0$. We should note that this disorder
GL-like theory is
valid strictly in $d=3$. Indeed, if we try to extrapolate to the
range of dimensionalities as in scaling relations of Section II,
we would obtain that the renormalized superfluid density, as
the ``charge'' of the DFT, scales like $\rho_s\sim\mu^{4-d}$, 
which would agree with Josephson's relation
{\it only} for $d=3$.

\subsection{Critical dynamics of the dual model}

In the duality transformations in the lattice only the 
phase of the order paramerer plays a role. \cite{Halperin,Kleinert} 
Actually, often duality arguments are worked out directly in the 
continuum, by relating the gradient of the phase to a 
``magnetic field'' that couples to the vortex loops within the 
dual model. \cite{Son,Wen} In the case of critical dynamics it is simpler 
to follow a similar approach in order to derive the main results.

The superfluid velocity ${\bf v}_s$ satisfies the dynamical equation 
\cite{Halperin}

\begin{equation}
\label{dynphase}
\frac{\partial{\bf v}_s}{\partial t}=\Gamma_0\rho_s^0\nabla^2{\bf v}_s+q{\bf E},
\end{equation} 
where $\Gamma_0$ is the bare kinetic coefficient, ${\bf E}$ is the 
electric field, and we have neglected the noise for simplicity. 
The superfluid velocity is related as usual to the phase of 
the order parameter, i.e., ${\bf v}_s=\mbox{\boldmath $\nabla$}\theta$. Because of 
the vortices, $\mbox{\boldmath $\nabla$}\times{\bf v}_s\neq 0$. The vortex 
current is given by

\begin{equation}
\label{vc}
{\bf J}_v=\frac{\sqrt{\rho_s^0}}{2\pi}~\mbox{\boldmath $\nabla$}\times{\bf v}_s.
\end{equation}
Thus, by taking the curl of Eq. (\ref{dynphase}), we obtain a dynamical 
equation for the vortex current:

\begin{equation}
\label{eq-vc}
\frac{\partial{\bf J}_v}{
\partial t}=\Gamma_0\rho_s^0\nabla^2{\bf J}_v-\frac{\mu_{A,0}}{2\pi}
\frac{\partial{\bf B}}{\partial t},
\end{equation}
where ${\bf B}$ is the macroscopic magnetic induction field. 
In Fourier space and at zero momentum we obtain

\begin{equation}
\label{vc-lr}
{\bf J}_v(\omega)=-\frac{\mu_{A,0}}{2\pi}{\bf B}(\omega).
\end{equation}
Thus, the above simple analysis already shows us that the 
introduction of dynamics in the duality approach leads to 
interesting consequences. In the original model linear response 
theory relates the current to the electric field in 
Fourier space as

\begin{equation}
\label{lin-resp}
{\bf J}(\omega,{\bf p})=\sigma(\omega,{\bf p}){\bf E}(\omega,{\bf p}).
\end{equation}
With respect to the dynamics of the system, the statistical mechanics 
duality implies also {\it electric-magnetic duality}, 
in which case the electric field is replaced by the (true) magnetic field 
in a linear response theory for the vortex current. 

Generally, the current of the original GL model is related to  
the vortex current in the dual model by the formula,  

\begin{equation}
\label{vortex-cur-dyn}
\mbox{\boldmath $\nabla$}\times{\bf J}(t,{\bf r})=
\int dt'\int d^3r' Q(t-t',{\bf r}-{\bf r}'){\bf J}_v(t',{\bf r}'),
\end{equation}
generalizing the classical static duality relation

\begin{equation}
\label{vortex-cur}
2\pi\mu_{A,0}~{\bf J}_v=\mbox{\boldmath $\nabla$}\times{\bf J},
\end{equation}
which is equivalent to Eq. (\ref{vc}), since ${\bf J}=q\rho_s^0{\bf v}_s$. 
Note that the factor $2\pi\sqrt{\rho_s^0}$ accounts for an elementary flux quantum in 
the dual model corresponding to a unit charge. \cite{Son} 
Eq. (\ref{vortex-cur}) also follows from the continuum limit 
of the exact duality transformation on the 
lattice. 

The low-frequency 
limit of $Q(\omega)\equiv Q(\omega,{\bf p}=0)$ is given by

\begin{equation}
Q(\omega)\approx 2\pi q\sqrt{\rho_s}=2\pi\mu_A.
\end{equation}
Thus, when the fluctuations are taken into account, Eq. (\ref{vortex-cur}) will hold 
approximately in the low-frequency limit with $\mu_{A,0}$ replaced by 
$\mu_A$. 

The electric-magnetic duality in the response 
functions implies a linear response

\begin{equation}
\label{lin-resp-dual}
{\bf J}_v(\omega,{\bf p})=\tilde{\sigma}(\omega,{\bf p}){\bf B}(\omega,{\bf p}),
\end{equation}
where $\tilde{\sigma}$ is the 
{\it dual AC conductivity}. The above equation is a 
generalization of Eq. (\ref{vc-lr}). 
In order to check the validity of 
Eq. (\ref{lin-resp-dual}), we rewrite Eq. (\ref{lin-resp}) in real 
space:

\begin{equation}
\label{lin-resp-rs}
{\bf J}(t,{\bf r})=\int dt'\int d^3r'\sigma(t-t',{\bf r}-{\bf r}')
{\bf E}(t',{\bf r}'),
\end{equation}
where ${\bf E}=-\partial{\bf A}/{\partial t}$. 
By taking the curl of  Eq. (\ref{lin-resp-rs}) we obtain, after some 
trivial algebra, precisely Eq. (\ref{lin-resp-dual}), with 
the dual conductivity related to the conductivity of the original 
model through the formula

\begin{equation}
\label{dualconduc}
\tilde{\sigma}(\omega)=\frac{i\omega}{2\pi\mu_A}
~\sigma(\omega), 
\end{equation}
where we have assumed a low frequency regime at zero momentum.

Since the DFT Hamiltonian (\ref{contdual}) has the same 
form as the GL model in the MCF regime, we have that 
an anomalous dimension $\eta_h=1$ is generated for the 
gauge field ${\bf h}$. Therefore, the scaling 
behavior of the AC conductivity in the MCF regime given by 
Eq. (\ref{conducscal1}) also applies here and we obtain 
that $\tilde{\sigma}(\omega)\sim\mu^{2-\tilde{z}}$, where $\tilde{z}$ 
denotes the dynamic critical exponent of the DFT. Thus, simple dimensional 
analysis in Eq. (\ref{dualconduc}), using the scaling relation 
(\ref{conducscal0}) for $d=3$, yields

\begin{equation}
\label{ztilde}
\tilde{z}=\frac{3}{2}.
\end{equation}
Note that the dynamic exponent $z$ drops out in the power counting. 
The above result would also follow by considering a renormalized 
version of Eq. (\ref{vc-lr}), where $\mu_{A,0}$ is replaced by 
$\mu_A$. In the extreme type II limit $\mu_A\sim\mu^{1/2}$, which 
should have the same scaling as $\tilde{\sigma}\sim\mu^{2-\tilde{z}}$, 
implying once more that $\tilde{z}=3/2$.   

The question now is how the exponent $z$ is related to 
the exponent $\tilde{z}$ of the dual model. We have seen that the static 
exponents coincide, for the obvious reason that both original and 
dual models lead to the same singular contribution for the 
free energy. The relation between $\tilde{z}$ and $z$ is much less 
obvious. It can be obtained as folows. From Eq. (\ref{ACconduccritXY}) and 
the Josephson relation we obtain that the right-hand side of Eq. (\ref{dualconduc}) 
scales for $d=3$ and ${\bf p}=0$ as

\begin{equation}
\label{dualconduc1}
\tilde{\sigma}(\omega)\sim(-i\omega)^{1/2z}.
\end{equation} 
Since $\tilde{\sigma}(\omega)$ must scale like in the MCF regime, we can use 
Eq. (\ref{ACconduccritMCF}) and write also 
\begin{equation}
\label{dualconduc2}
\tilde{\sigma}(\omega)\sim(-i\omega)^{(2-\tilde{z})/\tilde{z}}.
\end{equation}
Comparison of Eqs. (\ref{dualconduc1}) and (\ref{dualconduc2}) leads to the 
scaling relation

\begin{equation}
\label{scal-dual}
2z=\frac{\tilde{z}}{2-\tilde{z}}.
\end{equation}
Insertion of the exact result (\ref{ztilde}) in the above equation gives then the 
{\it exact} result

\begin{equation}
\label{z}
z=\frac{3}{2}.
\end{equation}

It is interesting to note that the scaling relation (\ref{scal-dual}) still gives a reasonable 
result for $\tilde{z}$ when the mean-field value $z=2$ is used, leading to 
$\tilde{z}=8/5=1.6$, which is not too discrepant with the exact value (\ref{ztilde}). 

The above results imply the following duality relation between the conductivities of the 
original and dual models, 

\begin{equation}
\label{dual-rel}
\sigma(\omega)\tilde{\sigma}(\omega)\sim{\rm const},
\end{equation}
as the critical point is approached. The above duality relation is analogous to  
the well known Dirac relation between electric and magnetic charges. \cite{Kleinert,Peskin} 

\section{Discussion}

In this paper we have combined RG and duality arguments in order to solve 
a controversial issue, namely, on the value of the dynamic critical exponent 
$z$ in extreme type II superconductors. Our analysis implies that 
$z=3/2$ exactly, therefore confirming the result given by the Monte 
Carlo simulations of Ref. \onlinecite{Lidmar}, and the 
experimental value of Ref. \onlinecite{Kim}. An interesting consequence 
of our analysis is that the original model and its dual share the 
same dynamic exponent. Thus, $z$, $\alpha$, and $\nu$ are the same in both 
models, while $\eta$ is not the same. \cite{Hove1} 

It remains to discuss how our analysis fits in the classification of 
dynamic models of Hohenberg and Halperin. \cite{Hohenberg} We have 
obtained a value of $z$ identical to the one of model E, which is a 
model critical dynamics for the spin-wave modes of a superfluid. Since 
the spin-waves are decoupled from the vortex degrees of freedom, \cite{Kleinert} and 
the duality analysis of the critical dynamics takes precisely the latter into 
account, model E should not be expected to govern the critical dynamics of an 
extreme type II superconductor at zero external field. The right model 
critical dynamics is actually given by a generalization of model A in which the 
dynamics of the vortex loops are taken into account. A simplified 
version of such a model was considered in the beginning of   
Section III-B, in Eqs. (\ref{dynphase}) and (\ref{eq-vc}). These equations 
describe respectively the dynamics of the supercurrent and the vortex 
current. An interesting physical consequence of such a dynamics is the 
electric-magnetic duality in the linear response of the supercurrent and 
the vortex current. Such a point of view leads to a modified dynamic London equation, 
which is derived from the two following Maxwell equations generalized in 
such a way to include also the vortex current,

\begin{equation}
\label{M1}
\mbox{\boldmath $\nabla$}\times{\bf E}=-\frac{\partial{\bf B}}{\partial t}-{\bf J}_v,
\end{equation}
\begin{equation}
\label{M2}
\mbox{\boldmath $\nabla$}\times{\bf B}={\bf J}+\frac{\partial{\bf E}}{\partial t}.
\end{equation}
The remaining Maxwell equations are not affected by the vortices, since 
$\mbox{\boldmath $\nabla$}\cdot{\bf J}_v=0$. Due to this, the electric-magnetic 
duality holds only in the above two Maxwell equations. It would only hold in all the 
four equations if open vortex lines were present, in which case magnetic 
monopoles would be attached to the vortex line ends. This is obviously not the case here 
since the $U(1)$ group is not compact. \cite{Smiseth} 

The electric-magnetic 
duality of the above equations 
corresponds to ${\bf E}\to{\bf B}$, ${\bf B}\to-{\bf E}$, and 
${\bf J}\to-{\bf J}_v$. We will derive the London equation using the 
approximation where ${\bf J}_v$ has a very weak dependence on ${\bf r}$, 
i.e., by using Eqs. (\ref{vc-lr}) and (\ref{vortex-cur}). Thus, by 
taking the curl of Eq. (\ref{M2} and taking into account 
Eqs. (\ref{vc-lr}), (\ref{vortex-cur}), and $\mbox{\boldmath $\nabla$}\cdot{\bf B}=0$, 
we obtain

\begin{equation}
\label{London-dyn}
\frac{\partial^2{\bf B}}{\partial t^2}-\nabla^2{\bf B}-
\mu_{A,0}^2{\bf B}+\frac{\mu_{A,0}}{2\pi}\frac{\partial {\bf B}}{\partial t}=0.
\end{equation}
Note that the above dynamic London equation contains a damping term. The general 
case is of course more complicate and Eqs. (\ref{lin-resp-dual}) and 
(\ref{dualconduc}) must be used instead. Remarkably, the duality approach allows us 
to derive an exact value for the dynamic exponent just by using scaling arguments. 
However, in order to derive the scaling functions for the AC conductivity a more 
explicit calculation is necessary. This will be the subject of a forthcoming 
work. \cite{Nogueira-inprep}  

We conclude by saying that although the dual dynamics studied here is different from 
the quantum case, \cite{Wen} the result (\ref{dual-rel}) applies also there.  
We should be note, however, that in the quantum case the 
dynamics can be derived from the quantum Lagrangian, which is not case for 
a classical system near the critical temperature.

\acknowledgments

The authors would like to thank W. Metzner and
H. Kleinert for helpful discussions.
F.S.N. would like to thank the Max-Planck-Institut 
f{\"u}r Festk{\"o}rperforschung, where this work was initiated, 
for the hospitality. F.S.N. also acknowledges the financial support 
of the DFG Priority Program SPP 1116 in earlier stages of this work.

\end{document}